\documentclass[aps,prl,twocolumn,nofootinbib,showpacs]{revtex4-1}

\usepackage{booktabs,makecell,multirow}
\usepackage{ulem}
\usepackage{graphicx,epsfig}
\usepackage{subfigure}
\usepackage{CJK}

\usepackage[colorlinks=true, linkcolor=blue, citecolor=blue, urlcolor=blue]{hyperref}

\begin{document}

\begin{CJK*}{GBK}{song}

\title{Novel and self-consistency analysis of the QCD running coupling $\alpha_s(Q)$ in both the perturbative and nonperturbative domains}

\author{Qing Yu $^{1,2}$}
%\email{yuq@cqu.edu.cn}

\author{Hua Zhou $^{1,2}$}
%\email{zhouhua@cqu.edu.cn}

\author{Xu-Dong Huang $^1$}
%\email{hxud@cqu.eud.cn}

\author{Jian-Ming Shen $^3$}
%\email{shenjm@hnu.edu.cn}

\author{Xing-Gang Wu $^{1}$}
%\email{wuxg@cqu.edu.cn}

\address{$^1$ Department of Physics, Chongqing Key Laboratory for Strongly Coupled Physics, Chongqing University, Chongqing 401331, People's Republic of China}
\address{$^2$ Department of Physics,
Norwegian University of Science and Technology, H{\o}gskoleringen 5,
N-7491 Trondheim, Norway}
\address{$^3$ School of Physics and Electronics, Hunan University, Changsha 410082, P.R. China}

\date{\today}

\begin{abstract}

The QCD coupling $\alpha_s$ is the most important parameter for achieving precise QCD predictions. By using the well measured effective coupling $\alpha^{g_1}_{s}(Q)$ defined from the Bjorken sum rules as a basis, we suggest a novel and self-consistency way to fix the $\alpha_s$ at all scales: The QCD light-front holographic model is adopted for its infrared behavior, and the fixed-order pQCD prediction under the principle of maximum conformality (PMC) is used for its high-energy behavior. Using the PMC scheme-and-scale independent perturbative series, and by transforming it into the one under the physical $V$-scheme, we observe that a precise $\alpha_s$ running behavior in both the perturbative and nonperturbative domains with a smooth transition from small to large scales can be achieved.


\end{abstract}

\maketitle

The QCD running coupling ($\alpha_s$) sets the strength of the interactions of quarks and gluons, whose correct and exact value is important for achieving precise QCD predictions. On the one hand, in the large scale (short-distance) region, due to the property of asymptotic freedom~\cite{Gross:1973id, Politzer:1973fx}, the magnitude of $\alpha_s$ becomes small and its scale-running behavior can be controlled by renormalization group equation (RGE). By using the RGE, one can fix its value at any large scale by using the measurements of the high-energy observables that fix $\alpha_s$ at a given scale. On the other hand, in small scale (long-distance) region, a natural extension of $\alpha_s$-behavior derived from the RGE shall meet the unphysical Landau singularity. Due to its perturbative nature, various theories and low-energy models have been suggested to set the $\alpha_s$ infrared behavior, cf. the reviews~\cite{Prosperi:2006hx, Deur:2016tte}. For example, the dilaton soft-wall modification of the ${\rm AdS}_5$ metric $e^{+\kappa^2 z^2}$ together with the QCD light-front holography (LFH)~\cite{Brodsky:2014yha}, where $\kappa$ is a confinement scale derived from hadron masses, predicts $\alpha_{s}/\pi\to 1$ for $Q^2\to 0$. It is helpful to find a proper way to fix the $\alpha_s$ value at all scales.

It has been suggested that one can define an effective QCD running coupling at all scales via a perturbatively calculable physical observable~\cite{Grunberg:1980ja, Grunberg:1982fw}. For example, by using the JLAB data on the Bjorken sum rules (BSR) $\Gamma^{p-n}_1(Q)$~\cite{Bjorken:1966jh, Bjorken:1969mm}, one can define an effective coupling $\alpha^{g_1}_s(Q)$ via the following way~\cite{Deur:2004ti, Deur:2005cf, Deur:2008ej, Deur:2014vea},
\begin{eqnarray}
\Gamma^{p-n}_1(Q)&=&\int^1_0 dx(g^p_1(x)-g^n_1(x))    \nonumber\\
&=&\frac{g_A}{6}\left[1-a^{g_1}_s(Q) \right],
\label{gammapn}
\end{eqnarray}
where $a^{g_1}_s(Q)= {\alpha^{g_1}_s(Q)} /{\pi}$ and $Q$ is the energy scale at which it is measured. $g_1^{p,n}(x)$ are spin structure functions for the proton and neutron with Bjorken scaling variable $x$, and $g_A$ is the nucleon axial charge. At the finite $Q^2$-range, the Bjorken sum rules $\Gamma^{p-n}_1(Q)$ is a generalized description of perturbative QCD (pQCD) corrections and non-perturbative power corrections. The non-perturbative corrections are usually parameterized as a series of over various powers of $1/Q^2$, which are highly suppressed in large $Q^2$-region. While in low and intermediate $Q^2$-regions, the non-perturbative terms shall have sizable contributions to $\Gamma^{p-n}_1(Q)$. A detailed discussion of non-perturbative contributions can found in~Ref.\cite{Yu:2021ofs}. To make the matching of $\alpha_s$ in perturbative and non-perturbative regions more transparent, the above so-defined effective coupling $a^{g_1}_s(Q)$ implicitly absorbs both the non-perturbative contributions and the higher-order perturbative contributions into the definition~\cite{Deur:2004ti, Deur:2005cf, Deur:2008ej, Deur:2014vea}. It thus provides a convenient platform for testing or fixing the running behavior of $\alpha_s$ at all scales.

At high momentum transfer, the effective coupling $a^{g_1}_s$ satisfies asymptotic freedom, which can be expanded as a series over the $\overline{\rm MS}$-scheme running coupling $a^{\overline{\rm MS}}_s$,
\begin{eqnarray} \label{conv.series}
a^{g_1}_s(Q)&=&\sum^{n}_{i=1}r^{\overline{\rm MS}}_{i}(Q, \mu_r) a^{\overline{\rm MS}, i}_s(\mu_r),
\end{eqnarray}
where $\mu_r$ is the renormaliztion scale and the perturbative coefficients $r_i$ have been calculated up to four-loop QCD corrections~\cite{Baikov:2010je, Baikov:2012zm}. Using this higher-loop pQCD series, we can achieve a precise prediction on $a^{g_1}_s(Q)$ at the high momentum transfer, and by requiring its value and its slope be matched to a low-energy model such as~\cite{Brodsky:2010ur}
\begin{eqnarray}
a^{ g_{1}, {\rm LFH}}_{s}(Q) = e^{-Q^2/4\kappa^2}.
\label{ag1LFH}
\end{eqnarray}
A comparison of various low-energy models can be found in Ref.\cite{Zhang:2014qqa}. Here we shall adopt the model (\ref{ag1LFH}) to do the matching, since its prediction agrees with the hadronic data extracted from various observables as well as the predictions of various models with the built-in confinement and lattice simulations. Some attempts have been done to fix an interface scale and a smooth connection between perturbative and non-perturbative hadron dynamics, cf. Refs.\cite{Brodsky:2010ur, Deur:2014qfa, Deur:2016cxb, Deur:2017cvd}.

In Refs.\cite{Brodsky:2010ur, Deur:2014qfa, Deur:2016cxb}, the scheme-and-scale dependent fixed-order pQCD series (\ref{conv.series}) has been adopted to do the matching, whose renormalization scale is set as the guessed typical momentum transfer of the process (e.g. $Q$) and an arbitrary range $[Q/2,2Q]$ is then assigned to estimate its uncertainty. Due to the mismatching of $\alpha_s$ and the coefficients at each perturbative order, this scale uncertainty is unavoidable, whose magnitude depends heavily on the how many terms of the pQCD series are known and the convergence of the pQCD series, and it is then conventionally treated as an important systematic error of the pQCD prediction. Numerically, it has been found that the scale errors are still sizable in intermediate and low-energy region even for the present known four-loop series due to larger $\alpha_s$ in those regions. This manly input and unwanted scale error thus greatly affects the accuracy of the matching. Thus it is important to adopt a proper scale-setting approach so as to achieve a more accurate fixed-order prediction.

In the literature, the principle of maximum conformality (PMC)~\cite{Brodsky:2011ta, Mojaza:2012mf, Brodsky:2011ig, Brodsky:2012rj, Brodsky:2013vpa} has been suggested to eliminate such scale errors. It uses the RGE and fixes the correct magnitude of $\alpha_s$ by absorbing all the $\{\beta_i\}$-related non-conformal terms via a systematic way, while remaining the scale-independent conformal coefficients. This leads to a scheme and scale invariant pQCD prediction~\cite{Wu:2014iba, Wu:2019mky}, which agrees well with the renormalization group invariance (RGI)~\cite{Petermann:1953wpa, GellMann:1954fq, Peterman:1978tb, Callan:1970yg, Symanzik:1970rt, Brodsky:1982gc}. In year 2017, the PMC multi-scale approach has been applied to do the matching of $a^{g_1}_s(Q)$ in both the perturbative and nonperturbative domains~\cite{Deur:2017cvd}. It has been found that a more precise matching of $\alpha^{g1}_s(Q)$ can be achieved, but they also met the ``self-consistency problem", i.e. the PMC scales at some orders are smaller than the critical scale $Q_0$, which represents the transition between the perturbative and non-perturbative QCD domains.

The lately suggested PMC single-scale approach~\cite{Shen:2017pdu, Wu:2018cmb} determines a single effective PMC scale by using the RGE, which represents the overall effective momentum flow of the process and replaces all the multi-scales at each order on the basis of a mean value theorem. The PMC single-scale approach is a reliable substitution for the PMC multi-scale approach, which also greatly suppress the residual scale dependence due to unknown perturbative terms~\cite{Huang:2021hzr}. The PMC predictions are scheme independent, which are ensured by the PMC conformal series, and by using the commensurate scale relations among different schemes~\cite{Brodsky:1994eh}, the determined PMC scale may be larger than the critical scale $Q_0$ by choosing a proper scheme other than the $\overline{\rm MS}$-scheme, then a solution of the previous ``self-consistency problem" may be achieved. After trying various intermediate schemes, we find that the physical $V$-scheme may achieve the goal. The $V$-scheme coupling $\alpha_s^V$ is gauge-independent and physical, which is defined in the static limit of the scattering potential between two heavy quark-antiquark test charges~\cite{Appelquist:1977tw, Fischler:1977yf, Peter:1996ig, Schroder:1998vy}
\begin{equation}
V(Q^2) = - 4 \pi C_F {\alpha^{V}_{s}(Q) \over Q^2}
\end{equation}
at the momentum transfer $q^2 = -Q^2$, where $C_F=({N_C^2-1})/({2N_C})$ is the Casimir operator for the fundamental representation of $SU(N_C)$-group with $N_C=3$ for QCD. The $V$-scheme coupling $\alpha_s^V$ has some advantages. It corrects the static potential by higher-order QCD corrections and is well-suited for summing the effects of gluon exchanges at low momentum transfer, such as in evaluating the final-state interaction corrections to heavy quark production~\cite{Brodsky:1995ds}, or in evaluating the hard-scattering matrix elements underlying the exclusive processes~\cite{Brodsky:1997dh}. Different from the $\overline{\rm MS}$-scheme, the $V$-scheme is also helpful to model a smooth transition of the QCD running coupling through the thresholds of heavy quark productions, since it corrects the massive dependent corrections in its running behavior~\cite{Brodsky:1998mf}.

In this letter, we shall show that the ``self-consistency problem" can indeed be solved by applying the PMC single-scale approach together with the use of the physical $V$-scheme. For the purpose, we first transform the known four-loop $\overline{\rm MS}$-scheme perturbative series (\ref{conv.series}) of $a^{g1}_s$ into the $V$-scheme one,
\begin{eqnarray}
a^{g_1}_s(Q)&=& r^{\rm V}_{1,0}a^{\rm V}_{s}(\mu_r) +(r^{\rm V}_{2,0}+\beta_0 r^{\rm V}_{2,1})a^{\rm V,2}_{s}(\mu_r) \nonumber\\
& & +(r^{\rm V}_{3,0}+\beta_1 r^{\rm V}_{2,1}+2\beta_{0} r^{\rm V}_{3,1}+\beta^2_{0} r^{\rm V}_{3,2}) a^{\rm V,3}_{s}(\mu_r) \nonumber\\
& & +(r^{\rm V}_{4,0}+\beta^{\rm V}_2 r^{\rm V}_{2,1} +2\beta_{1} r^{\rm V}_{3,1} + \frac{5}{2} \beta_0 \beta_1 r^{\rm V}_{3,2} \nonumber\\
& & +3\beta_0 r^{\rm V}_{4,1} +3 \beta^2_0 r^{\rm V}_{4,2} + \beta^3_0 r^{\rm V}_{4,3}) a^{\rm V,4}_{s}(\mu_r) ,
\label{ag1con}
\end{eqnarray}
where the QCD degeneracy relations~\cite{Bi:2015wea} have been implicitly adopted to transform $r^{\overline{\rm MS}}_{i}$ into $r^{\rm V}_{i,j}$ and the $V$-scheme $\{\beta^{\rm V}_i\}$-functions can be derived by using their relations to the $\overline{\rm MS}$-scheme ones~\cite{Chetyrkin:2004mf, Czakon:2004bu, Baikov:2016tgj}, $\beta^{\rm V}(a_s^{\rm V}) =\left(\partial a_s^{\rm V} / \partial a_s^{\rm \overline{MS}}\right) \beta^{\rm \overline{MS}} (a_s^{\rm \overline{MS}})$. The coefficients $r^{\rm V}_{i,j}~(j\neq0)$ are general functions of $\ln({\mu^2_r}/{ Q^2})$; i.e.
\begin{equation}
r^{\rm V}_{i,j} = \sum_{k=0}^{j} C_j^k \ln^k(\mu_r^2/Q^2) \hat{r}^{\rm V}_{i-k,j-k},  \label{rij}
\end{equation}
where the coefficients $C_j^k={j!}/{k!(j-k)!}$ and the coefficients $\hat{r}^{\rm V}_{i,j}=r^{\rm V}_{i,j}|_{\mu_r=Q}$. The magnitude of $\alpha_s$ can be determined by using the $\{\beta^{\rm V}_i\}$-functions.

Following the standard PMC procedures, by requiring all the RGE-involved $\{\beta_i\}$-terms to zero, one can determine a scale-invariant optimal scale $Q^{\rm V}_\star$ of the process and obtain a conformal series as follows
\begin{equation}
{a^{g_1}_{s}}(Q)|_{\rm PMC} = \sum^{4}_{i\ge1} \hat{r}^{\rm V}_{i,0} a^{\rm V,i}_s(Q_\star),
\label{ag1pmc}
\end{equation}
where the scale $Q_{\star}$ is a function of $Q$, which by using the four-loop pQCD series, can be fixed up to next-to-next-to-leading-log (NNLL) accuracy,
\begin{eqnarray}
\ln{\frac{Q^{2}_{\star}}{Q^2}}&=&  T_{0} + T_{1} a^{\rm V}_{s}(Q) + T_{2} a^{\rm V,2}_{s}(Q),
\label{ag1pmcscale}
\end{eqnarray}
where
\begin{eqnarray}
T_0 &=& -{\hat{r}^{\rm V}_{2,1}\over \hat{r}^{\rm V}_{1,0}},\\
T_1 &=& {2(\hat{r}^{\rm V}_{2,0}\hat{r}^{\rm V}_{2,1}-\hat{r}^{\rm V}_{1,0}\hat{r}^{\rm V}_{3,1})\over \hat{r}^{\rm V,2}_{1,0}}  +{(\hat{r}^{\rm V,2}_{2,1}-\hat{r}^{\rm V}_{1,0}\hat{r}^{\rm V}_{3,2})\over \hat{r}^{\rm V,2}_{1,0}}\beta_0,\\
T_{2}&=&\frac{4 (\hat{r}^{\rm V}_{1,0}\hat{r}^{\rm V}_{2,0}\hat{r}^{\rm V}_{3,1}-\hat{r}^{\rm V,2}_{2,0}\hat{r}^{\rm V}_{2,1})
+3(\hat{r}^{\rm V}_{1,0}\hat{r}^{\rm V}_{2,1}\hat{r}^{\rm V}_{3,0}-\hat{r}^{\rm V,2}_{1,0}\hat{r}^{\rm V}_{4,1})}{ \hat{r}^{\rm V,3}_{1,0}}\nonumber\\
&&+\frac{3(\hat{r}^{\rm V,2}_{2,1}-\hat{r}^{\rm V}_{1,0}\hat{r}^{\rm V}_{3,2})}{2\hat{r}^{\rm V,2}_{1,0}}
\beta_1 -\frac{\hat{r}^{\rm V}_{2,0}\hat{r}^{\rm V,2}_{2,1} +3\hat{r}^{\rm V,2}_{1,0}\hat{r}^{\rm V}_{4,2}}{\hat{r}^{\rm V,3}_{1,0}}\beta_0\nonumber\\
&&-\frac{2(\hat{r}^{\rm V}_{2,0}\hat{r}^{\rm V,2}_{2,1}-
2\hat{r}^{\rm V}_{1,0}\hat{r}^{\rm V}_{2,1}\hat{r}^{\rm V}_{3,1}-\hat{r}^{\rm V}_{1,0}\hat{r}^{\rm V}_{2,0}\hat{r}^{\rm V}_{3,2})}{\hat{r}^{\rm V,3}_{1,0}}\beta_0\nonumber\\
&&+\frac{(2\hat{r}^{\rm V}_{1,0}\hat{r}^{\rm V}_{2,1}\hat{r}^{\rm V}_{3,2}-\hat{r}^{\rm V,2}_{1,0}\hat{r}^{\rm V}_{4,3}-\hat{r}^{\rm V,3}_{2,1})}{\hat{r}^{\rm V,3}_{1,0}}\beta^{2}_0 .
\end{eqnarray}
It is noted that the perturbative series of $a^{g_1}_s(Q)|_{\rm{PMC}}$ is explicitly free of $\mu_r$, leading to a precise scheme-and-scale invariant fixed-order prediction. Due to the elimination of divergent renormalon terms, the convergence of the pQCD series can also be greatly improved.

\begin{figure}[htb]
\includegraphics[width=0.45\textwidth]{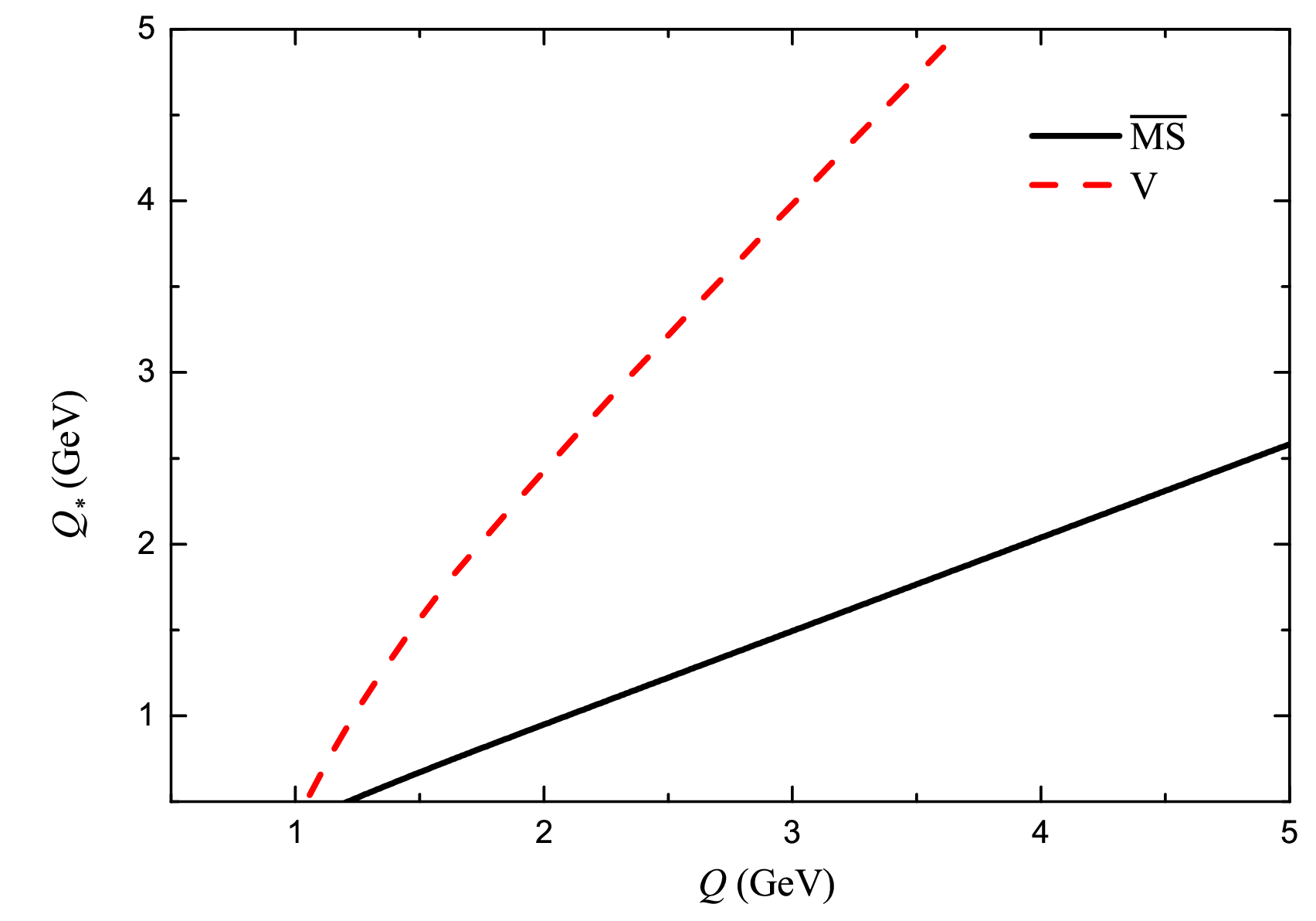}
\caption{The calculated effective scale $Q_\star$ up to NNLL accuracy under $\overline{\rm MS}$ scheme and V scheme, respectively.}
\label{Qstar}
\end{figure}

To do the numerical calculation, we adopt $\alpha_s(M_Z)=0.1179\pm0.0010$~\cite{PDG:2020} to fix the QCD asymptotic scale $\Lambda$, and we obtain $\Lambda^{\overline{\rm MS}}_3=0.343\pm{0.015}$ GeV and $\Lambda^{\rm V}_3=0.438\pm{0.019}$ GeV for three active flavors. As a typical example, the perturbative series (\ref{ag1pmc}) for $n_f=3$ becomes
\begin{eqnarray}
{a^{g_1}_{s}}(Q)|_{\rm PMC} &=& a^{\rm V}_s (Q_\star)+3.15 a^{\rm V,2}_s (Q_\star) \nonumber\\
& & + 20.46 a^{\rm V,3}_s(Q_\star)+51.36 a^{\rm V,4}_s (Q_\star),
\label{ag1v}
\end{eqnarray}
where $Q_\star$ satisfies
\begin{eqnarray}
\ln\frac{{Q^{2}_{\star}}}{Q^2} &=& 0.58+2.06\alpha^{\rm V}_s(Q)-7.41\alpha^{\rm V,2}_s(Q),
\label{pmcscalev}
\end{eqnarray}
which leads to $Q_{\star}=3.98$ GeV for $Q=3~{\rm GeV}$. Fig.~\ref{Qstar} shows how $Q_\star$ changes with $Q$, where $Q_\star$ under $\overline{\rm MS}$-scheme is also presented as a comparison. Fig.~\ref{Qstar} shows that $Q_\star$ under $V$-scheme has a faster increasing behavior with the increment of $Q$ than that of $\overline{\rm MS}$-scheme. Thus the previous puzzle of $Q_\star < Q_0$ can be solved. $\ln{Q^{2}_{\star}}/{Q^2}$ is a perturbative series, its unknown perturbative terms shall lead to \textit{the first kind of residual scale dependence}~\cite{Zheng:2013uja}. Since Eq.(\ref{pmcscalev}) already shows a perturbative behavior in large $Q^2$-region, as a conservative estimation, the magnitude of the N$^3$LL-terms can be taken as the NNLL one, e.g. $\pm 7.41\alpha^{\rm V,2}_s(Q)$, which gives $\Delta Q_\star\simeq \left(_{-1.05}^{+1.42}\right)~{\rm GeV}$ for $Q=3~{\rm GeV}$ (By using the PAA, the predicted $\Delta Q_\star$ is similar, which shall also be constrained by the matching criteria.). Numerically, we observe that such scale uncertainty shall be further constrained by the matching of $\alpha^{g_1}_s(Q)$ in perturbative and non-perturbative domains and we finally have $\Delta Q_\star(Q=3{\rm GeV})\simeq \left(_{-0.24}^{+1.10}\right)~{\rm GeV}$.

Similarly, the unknown higher-order terms of Eq.(\ref{ag1v}) shall lead to \textit{the second kind of residual scale dependence}~\cite{Huang:2021kzc}, which can be estimated by using a more strict Pad$\acute{e}$ approximation approach (PAA) due to more loop terms have been known~\cite{Basdevant:1972fe}. The PAA offers a feasible conjecture that yields the $5_{\rm th}$-order terms from the given $4_{th}$-order perturbative series, and a $[N/M]$-type approximant $\rho_4(Q)$ for ${a^{g_1}_{s}}(Q)|_{\rm PMC}$ is defined as
\begin{eqnarray}
\rho^{[N/M]}_4(Q)  &=&  a_s^{\rm V}(Q_\star) \frac{b_0+b_1 a_s^{\rm V}(Q_\star) + \cdots + b_N a_s^{{\rm V},N}(Q_\star)} {1 + c_1 a_s^{\rm V}(Q_\star) + \cdots + c_M a_s^{{\rm V},M}(Q_\star)} \nonumber \\
                    &=& \sum_{i=1}^{4} \hat{r}^{\rm V}_{i,0} a_s^{V,i}(Q_\star) + \hat{r}^{\rm V}_{5,0}\; a_s^{V,5}(Q_\star) +\cdots,
\end{eqnarray}
where the parameter $M\geq 1$ and $N+M=3$. The known coefficients $ \hat{r}^{\rm V}_{i\leq4,0}$ determine the parameters $b_{i\in[0,N]}$ and $c_{j\in[1,M]}$, which inversely predicts a reasonable value for the uncalculated ${\rm N^4LO}$-coefficient $\hat{r}^{\rm V}_{5,0}$~\cite{Du:2018dma}, i.e.
\begin{eqnarray}
{\hat r}_{5,0}^{\rm PAA} &=& \frac{\hat{r}^4_{2,0}-3\hat{r}_{1,0}\hat{r}^2_{2,0}\hat{r}_{3,0} +\hat{r}^2_{1,0}\hat{r}^2_{3,0}+2\hat{r}^2_{1,0}\hat{r}_{2,0} \hat{r}_{4,0}}{\hat{r}^3_{1,0}}.
\label{PAA03}
\end{eqnarray}
Then the uncertainty from the unknown terms could be estimated by $\pm {\hat r}_{5,0}^{\rm PAA}a^{\rm V,5}_s(Q_\star) =\pm 232.22 a^{\rm V,5}_s(Q_\star)$.

\begin{figure}[htb]
\centering
\includegraphics[width=0.45\textwidth]{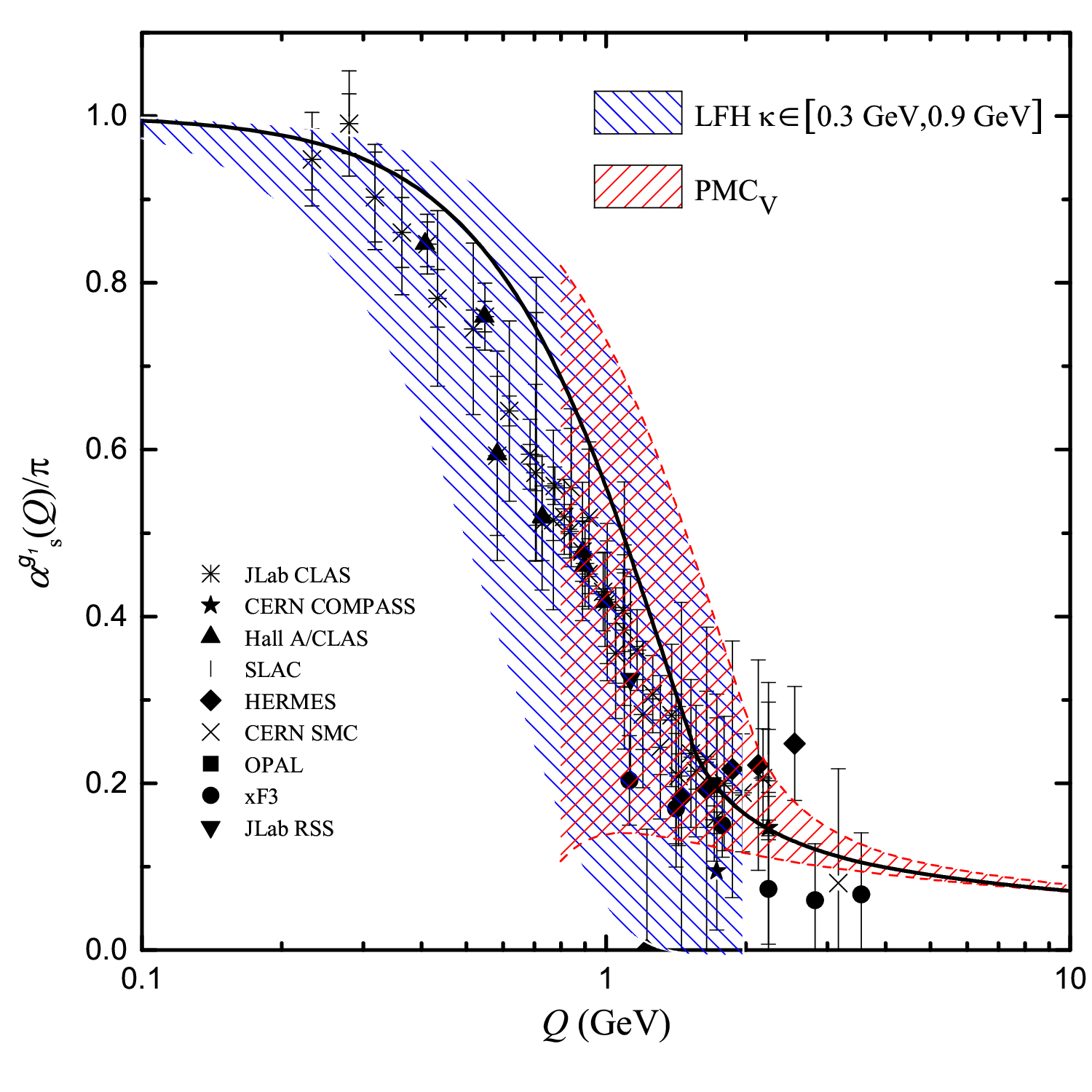}
\caption{The LFH low-energy $a^{g_1, {\rm LFH}}_s(Q)$ and the PMC prediction of $a^{g_1}_s(Q)|_{\rm PMC}$ under the $V$-scheme up to the ${\rm N^3LO}$-order QCD corrections. The blue band is the LFH model predictions $a^{g_1, {\rm LFH}}_s(Q)$ when taking parameter $\kappa$ in $[0.3~\rm GeV,0.9~\rm GeV]$. The red band is the uncertainty caused by squared averages of the residual scale dependence due to the uncalculated higher-order terms and $\Delta\alpha_s(M_Z)=\pm0.0010$. }
\label{vfig1}
\end{figure}

\begin{figure}[htb]
\centering
\includegraphics[width=0.45\textwidth]{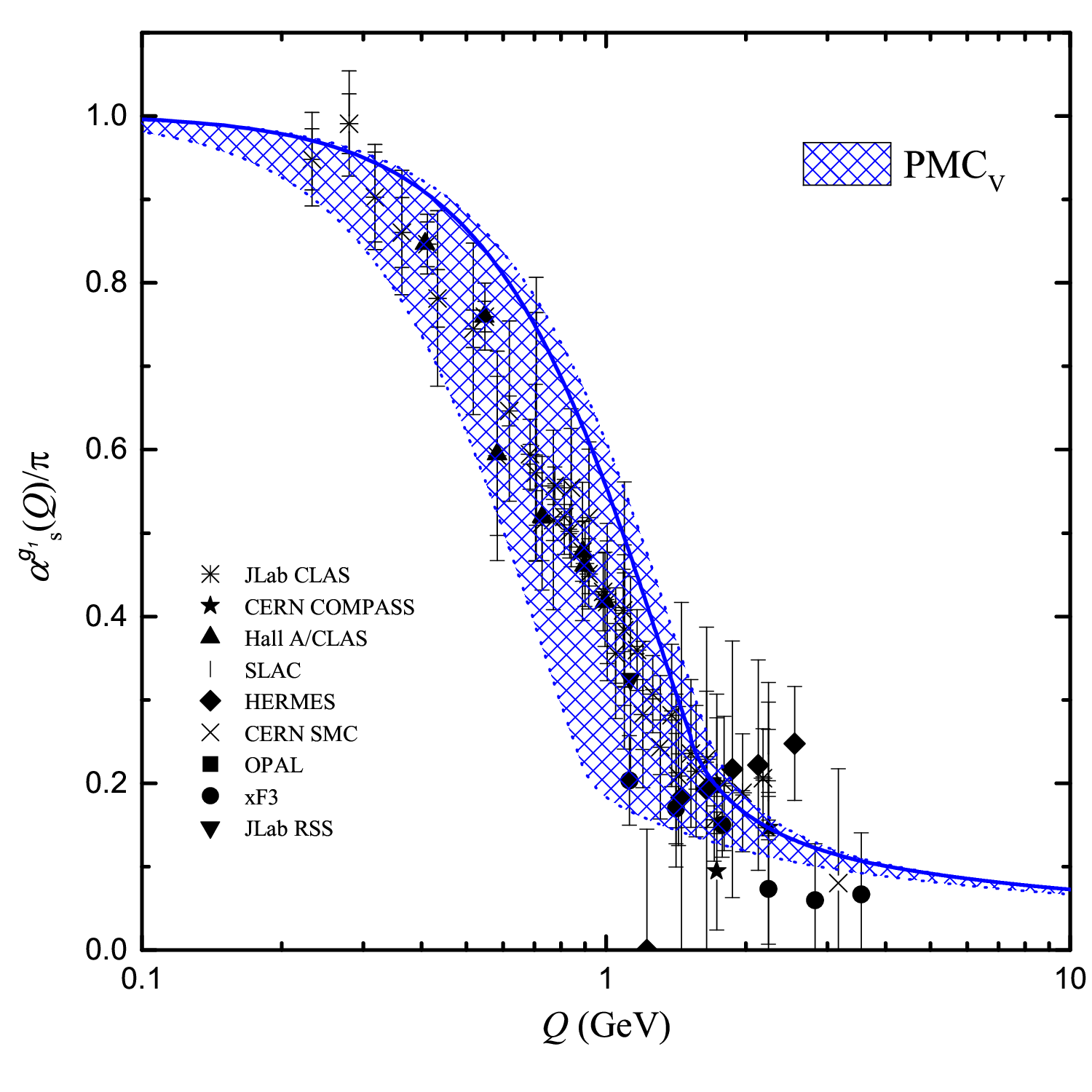}
\caption{The matching of the LFH low-energy $a^{g_1, {\rm LFH}}_s(Q)$ and the PMC prediction of $a^{g_1}_s(Q)|_{\rm PMC}$ under the $V$-scheme up to the ${\rm N^3LO}$-order QCD corrections. The shaded band is the uncertainty caused by squared averages of the residual scale dependence due to the uncalculated higher-order terms and $\Delta\alpha_s(M_Z)=\pm0.0010$. }
\label{vfig}
\end{figure}

It is found that the LFH model $a^{g_1, {\rm LFH}}_{s}(Q)$ can be naturally matched to the conformal perturbative series, since it is consistent with the conformal behavior at $Q^2\to 0$. To do the matching, we require the magnitudes and the derivatives of both the LFH $a^{g1, {\rm LFH}}_{s}(Q)$ and the prediction ${a^{g_1}_{s}}(Q)|_{\rm PMC}$ to be the same at the critical scale $Q_0$. We present $a^{g_1, {\rm LFH}}_s(Q)$ and the $V$-scheme $a^{g_1}_s(Q)|_{\rm PMC}$ up to ${\rm N^3LO}$-order QCD corrections in Fig.~\ref{vfig1}, where the LFH $a^{g_1, {\rm LFH}}_{s}(Q)$ is drawn by varying $\kappa$ within a wide range $[0.3~{\rm GeV},0.9~{\rm GeV}]$ and the available data issued by various experimental groups~\cite{Ackerstaff:1997ws, Alexakhin:2006oza, Anthony:1993uf, Abe:1997cx, Anthony:1999py, Adeva:1998vv, Ackerstaff:1998yj, Brodsky:2002nb, Deur:2004ti, Gross:1969jf, Kim:1998kia, Adeva:1998vv} have also been presented.  Fig.~\ref{vfig1} shows that to achieve a smooth connection, not all magnitudes of the conservatively estimated N$^3$LL-terms in $\ln{Q^{2}_{\star}}/{Q^2}$ are accepted; and by taking $\approx\left(^{+5.93}_{-1.48}\right)\alpha_s^{V,2}(Q)$, a well matching can be achieved, which is shown in Fig.~\ref{vfig}. The determined critical scale $Q_0=1.51^{+0.16}_{-0.62}$~GeV, whose errors are caused by \textit{the first kind of residual scale dependence} and \textit{the second kind of residual scale dependence} together with the error of $\Delta\alpha_s(M_Z)=\pm0.0010$. The input parameter of the LHF model $\kappa$ is $0.64^{+0.07}_{-0.28}$~GeV (This value is slightly larger than $\sim 1/2$ GeV, which incorporates high-twist contributions, since the high twist terms could have sizable contributions to $\Gamma^{p-n}_1(Q)$ in low-energy region~\cite{Yu:2021ofs}.), and the PMC scale $Q_{\star}(Q_0)\simeq 1.58^{+0.09}_{+0.02}$~GeV. We observe that $Q_{\star}(Q_0)$ is always larger than $Q_0$, thus the previous ``self-consistency problem"  is solved by using the $V$-scheme and the PMC single-scale approach.

Moreover, the quality of fit for the matched $a^{\rm g_1}_s$ can be measured by using the parameter $\chi^2/d.o.f$~\cite{PDG:2020}, which represents the quality $\chi^2$ over the number of experiment data points $N$ and is defined as
\begin{eqnarray}
\chi^2/d.o.f &=& \frac{1}{N-d}\sum^N_{j=1} \frac{1}{\sigma^2_{j, {\rm exp.}}+\sigma^2_{j,{\rm the.}}} \times  \nonumber\\
&& \quad\quad\quad\quad\quad \bigg[a^{g_1, {\rm exp.}}_{s}(Q_j)-a^{g_1, {\rm the.}}_{g1}(Q_j)\bigg]^2,
\end{eqnarray}
where ``the." stands for theoretical prediction and ``exp." stands for the experimental value, $\sigma^2_{j, {\rm exp.}}$ stands for the squared sum of the statistical error and systematic error at each data point $Q_j$. We adopt $N=70$, which are given in Refs.\cite{Ackerstaff:1997ws, Alexakhin:2006oza, Anthony:1993uf, Abe:1997cx, Anthony:1999py, Adeva:1998vv, Ackerstaff:1998yj, Brodsky:2002nb, Deur:2004ti, Gross:1969jf, Kim:1998kia, Adeva:1998vv}, and $d = 2$ due to two input parameters ($\kappa$ and $Q_0$). Our numerical calculation shows $\chi^2/d.o.f \simeq 0.12$, which corresponds to $p\simeq 99\%$, indicating a good goodness-of-fit  and the reasonableness of the fitted two input parameters.

As a final remark, we also calculate the correlation coefficient $\rho_{XY}$~\cite{PDG:2020} to show to what degree the matched $\alpha^{\rm g_1}_{s}$ are correlated to those $70$ data points
\begin{eqnarray}
\rho_{\rm XY}&=&\frac{\rm Cov(X,Y)}{\sigma_{\rm X}\sigma_{\rm Y}},
\end{eqnarray}
where $X$ and $Y$ stand for the experimental data on $a^{g_1}_s$ and the theoretically predicted ones, respectively. The covariance ${\rm Cov(X,Y)}=E[({\rm X}-E({\rm X}))({\rm Y}-E({\rm Y}))]=E({\rm XY})-E({\rm X})E({\rm Y})$, where $E({\rm X})$ stands the expectation value of $X$, $\sigma_{X,Y}$ represents the standard deviations of $X$ or $Y$. Numerically, we obtain $\rho_{\rm XY}\sim 0.96^{+0.02}_{-0.03}$, which indicates a high consistency between the predicted $a^{g_1}_s$ and the measured one.

{\bf\normalsize Summary.} The QCD running coupling is one of the most important parameter for QCD theory. By using the effective coupling $\alpha^{\rm g1}_{s}(Q)$ as an example, we have shown that a self-consistency $\alpha_s(Q)$ in both the perturbative and non-perturbative domains can be achieved by applying the PMC singlet-scale approach. Though the PMC prediction is scheme independent, a proper choice of scheme could have some subtle differences. Fig.~\ref{Qstar} shows that the effective PMC scale $Q_\star$ under the $V$-scheme has a faster increasing behavior with the increment of $Q$ than that of $\overline{\rm MS}$-scheme. Thus the previous puzzle of $Q_\star < Q_0$ is solved. The PMC eliminates the conventional renormalization scale ambiguity, and its single-scale setting approach greatly depresses the residual scale dependence due to uncalculated terms, thus achieving a more precise fixed-order pQCD prediction. We observe that the LFH low-energy model $a^{g_1, {\rm LFH}}_{s}(Q)$ can be naturally matched to the PMC conformal perturbative series ${a^{g_1}_{s}}(Q)|_{\rm PMC}$ over the physical $V$-scheme, and as shown by Fig.~\ref{vfig}, one can achieve a reasonable and smooth connection between the perturbative and non-perturbative domains.

\begin{acknowledgments}
We thank Stanley J. Brodsky for helpful discussions. This work was supported in part by the Chongqing Graduate Research and Innovation Foundation under Grant No.CYB21045 and No.ydstd1912, by the Natural Science Foundation of China under Grant No.11905056, No.12175025, and No.12147102, and by the Fundamental Research Funds for the Central Universities under Grant No.2020CQJQY-Z003.
\end{acknowledgments}

\end{CJK*}

\end{document}